\documentclass[smallextended]{svjour3} 
\smartqed  

\usepackage{graphicx} 

\newcommand{\kms}{km\,s$^{-1}$}
\newcommand{\arcsec}{$^{\prime\prime}$}

\begin{document}


\title{The co-evolution of galaxies and supermassive black holes in the near Universe}

\author{Thaisa Storchi-Bergmann}
\institute{Instituto de F'sica, Universidade Federal do Rio Grande do Sul \\
 IF--UFRGS, CP 15051, 91501-970, Porto
Alegre, RS, Brazil\\
\email{thaisa@ufrgs.br}}

\date{Received: date / Accepted: date}

\maketitle

\begin{abstract} 

A fundamental role is attributed to supermassive black holes (SMBH), and the feedback they generate,
in the evolution of galaxies. But theoretical models trying to reproduce the M$_{SMBH}-\sigma$ relation
(between the SMBH mass and stellar velocity dispersion of the galaxy bulge) make broad assumptions about the physical processes involved. These assumptions are needed due to the scarcity of observational constraints on the relevant physical processes which occur  when the SMBH is being fed via mass accretion in Active Galactic Nuclei (AGN). In search for these constraints, our group -- {\it AGN Integral Field Spectroscopy (AGNIFS)} -- has been mapping the gas kinematics as well as the stellar population properties of the inner few hundred parsecs of a sample of nearby AGN hosts.
In this contribution, I report results obtained so far which show gas inflows along nuclear spirals and compact disks in the inner tens to hundreds of pc in nearby AGN hosts which seem to be the sources of fuel to the AGN. As the inflow rates are much larger than the AGN accretion rate, the excess gas must be depleted via formation of new stars in the bulge.  Indeed, in many cases, we find $\sim$\,100\,pc circumnuclear rings of recent  star formation (ages\,$\sim$\,10--500\,Myr) that  can be interpreted as a signature of co-evolution of the host galaxy and its AGN. I also report the mapping of outflows in ionized gas, which are ubiquitous in Seyfert galaxies, and discuss mass outflow rates and powers. 


\end{abstract}


\section{Introduction}

The correlation between the mass of the Supermassive Black Hole (hereafter SMBH) present  in the nuclei of most galaxies -- the so-called M$_{SMBH}-\sigma$ relation -- has been interpreted as indicating a coupling between the growth of SMBH and that of  their host galaxies  \cite{ff05}.  As pointed out by Tiziana di Matteo in this conference (see also \cite{dimatteo05,dimatteo08}), the bulk of galaxy and  SMBH growth must have taken place in the first 1--2\,Gyr of the Universe. The observation of gas rich and star-forming galaxies, as well as the nuclear activity in Quasars at redshifts  z$\ge$6 support this conclusion and have motivated the cosmological simulations which show that  interactions can send gas towards the galaxy centers triggering episodes of star formation as well as luminous nuclear activity. 

Feeding and feedback processes which occur in Active Galactic Nuclei (AGN) are now a paradigm of galaxy evolution models in constraining the co-evolution of galaxies and SMBHs, but their implementation have been simplistic \cite{springel05,croton06,somerville08} because the physical processes involved are not well constrained by observations. This is due to the fact that  they occur within the inner few hundred parsecs, which cannot be  spatially resolved at $z\ge2$ where the co-evolution of galaxies and SMBH largely occurs. It is nearby galaxies that offer the only opportunity to test in detail the prescriptions used in models of galaxy and BH co-evolution.



In this contribution, I discuss results of recent studies of the gas kinematics within the inner few hundred parsecs of nearby active galaxies performed by my research group -- called  {\it AGNIFS (AGN Integral Field Spectroscopy)}, at spatial resolution of a few to tens of parsecs, which do resolve the gas kinematics in the nuclear region and can be used to constrain the processes of AGN feeding and feedback. In a few cases, we have been also able to map the stellar population in the nuclear region, in search of recent episodes of star formation which could trace the growth of the galaxy bulge.



\section{Observations}

We have used integral field spectroscopy (IFS) at the Gemini telescopes, the final product of which are ``datacubes'', which have two spatial dimensions --- allowing the extraction of images at a desired wavelength range ---
and one spectral dimension --- allowing the extraction of spectral information of each spatial element.

In the optical, we have used the Integral Field Unit of the Gemini Multi-Object Spectrograph (IFU-GMOS), which has a field-of-view of $3.\!''5\ \times 5$\arcsec\ in one-slit mode or 5\arcsec\,$\times$\,7\arcsec\ in two-slit mode at a sampling of $0.\!''2$ and angular resolution (dictated by the seeing) of $0.\!''6$, on average. The resolving power is $R \approx 2500$.

In the near-infrared (near-IR) we have used the Near-Infrared Integral Field Spectrograph (NIFS) together with the adaptative optics module ALTAIR (ALTtitude conjugate Adaptive optics for the InfraRed), which delivers an angular resolution of $\sim\!0.\!''1$. The field-of-view is 3\arcsec\,$\times$\,3\arcsec\ at a sampling of $0.\!''04 \times 0.\!''1$ and the spectral resolution is $R\approx 5300$ at the Z, J, H and K bands.

\section{Inflows}

Nuclear spirals --- on scales of hundred parsecs --- are frequently observed around AGN in images obtained with the {\em Hubble Space Telescope (HST)} \cite{martini03}. \cite{martini99} have shown that these spirals may be the channels through which matter is being transferred to the nucleus to feed the AGN. This interpretation is supported by models \cite{maciejewski04} and by results such as those from \cite{sl07}. In the latter paper, we have built ``structure maps'' using images obtained with the {\em HST} Wide-Field and Planetary Camera 2 (WFPC2) through the F606W filter of a sample of AGN and a control sample of non-active galaxies. The structure maps revealed dusty nuclear spirals in all early-type AGN hosts, but in only $\sim$\,25\% of the non-AGN, indicating that these spirals are strongly linked to the nuclear activity and thus should map the matter in its way to feed the SMBH at the nucleus. But in order to test this hypothesis, based only on morphology, it was necessary to map the gas kinematics in these spirals, what can be best done with IFUs.

\subsection{IFU observations in the optical}

\begin{figure}
\includegraphics[scale=0.8]{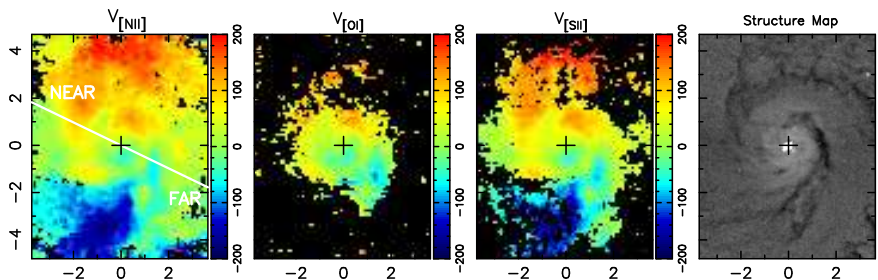}
\caption{First three panels: gas kinematics obtained from the centroid of the [N\,II]$\lambda$6584\AA, [O\,I]$\lambda$6300\AA\ and [S\,II]$\lambda$6717\AA\  emission lines of the inner  0.7\,kpc$\times$\,1\,kpc of NGC\,7213. Right panel: structure map of an HST F606W image of the same region. White line shows the line of nodes, which runs approximately from North (left) to South (right). West is to the top and right of the line of nodes \cite{sm13}.}
\label{velgas}
\end{figure}

We show in Fig.\,\ref{velgas} recent observations and measurements of the stellar and gas kinematics of the inner 0.7\,kpc$\times$1\,kpc of the LINER/Seyfert\,1 galaxy NGC\,7213, from \cite{sm13}. The observations, obtained with the GMOS IFU, cover the wavelength range 5700\AA--6900\AA. 

Although the galaxy appears to be close to face on, the stellar kinematics, obtained from the absorption lines, shows a rotation pattern with an amplitude of $\approx$\,50\,\kms, with the line of nodes oriented approximately along North-South (white line in Fig.\,\ref{velgas}). The gas kinematics -- obtained over the whole field-of-view in the [N\,II]$\lambda$6584\AA\ emission line -- is nevertheless completely distinct, showing a much larger amplitude (velocities up to 200\,\kms), and a ``distorted" rotation pattern with the largest velocity gradient running at a large angle to the line of nodes of the stellar kinematics. The ``distortions" in the gas velocity field are clearly correlated with the nuclear spirals seen in the structure map (shown in the rightmost panel of Fig.\,\ref{velgas}).  Considering that the near side of the galaxy is the West (to the top and right of the line of nodes in the first panel of Fig.\,\ref{velgas}), and the far side is the East, and assuming that the emitting gas is in the plane of the galaxy, the redshifts observed to the West and the blueshifts observed to the East can be interpreted as due to gas inflows towards the central region of the galaxy. In order to obtain an estimate for the mass inflow rate, we have integrated the mass flux through concentric ``shallow" cylinders around the nucleus, obtaining a mass inflow rate at a distance of $\approx$\,100\,pc from the nucleus of 0.2\,M$_\odot$\,yr$^{-1}$.

Similar observations and velocity fields were obtained for three other LINER galaxies. In M\,81 (LINER/Seyfert 1) we  \cite{sm11} have also observed rotation in the stellar velocity field within the inner 100\,pc radius, but a totally distinct kinematics for the gas. The ionized gas kinematics show inflows along the galaxy minor axis that seem to be correlated with a nuclear spiral seen in a structure map. The estimated ionized gas mass inflow rate in M\,81 is smaller than that in NGC\,7213, being of the order of the AGN accretion rate. Signatures of inflows along nuclear spirals at similar mass inflow rates as that obtained for M\,81 were also seen in the LINER/Seyfert\,1 galaxy NGC\,1097 \cite{fathi06} and in the LINER galaxy NGC\,6951 \cite{sb07}. We note that these inflows were obtained only from observations of the ionized gas, and thus can be considered a lower limit for the total mass gas inflow rate, which is probably dominated by cold molecular gas with possible contribution of neutral gas as well.

\subsection{IFU observations in the near-IR}

In \cite{riffel08} we have obtained the stellar and molecular gas kinematics of the inner 100\,pc of the nearby Seyfert 1 galaxy NGC\,4051 at a spatial resolution of 5\,pc, via a datacube obtained with NIFS in the near-IR  K band. The stellar kinematics shows again rotation with the lines of nodes running approximately North-South. As the spectral resolution of NIFS is almost three times higher than that of the GMOS IFU,  we  were able to obtain the gas kinematics using channel maps along the H$_2\lambda2.122\mu$m emission line. We found mostly blueshifts in a spiral arm to the East,  -- the far side of the galaxy, and mostly redshifts to the West -- the near side of the galaxy, which can be interpreted as inflows towards the galaxy center if we assume again that the gas is in the plane of the galaxy.

The H$_2$ emission in the K band maps the ``warm" molecular gas (temperature T$\sim$2000K) which is probably only the ``skin" of  a much larger cold molecular gas reservoir which emits in the millimetric wavelength range (and should thus be observable with the {\it Atacame Large Millimetric Array -- ALMA}, for example). In fact, the warm H$_2$ mass inflow rate within the inner 100\,pc of NGC\,4051 is of the order of only 10$^{-5}$\,M$_\odot$\,yr$^{-1}$. But previous observations of both the warm and cold  molecular gas in a sample of AGN host galaxies show  typical ratios cold/warm H$_2$ masses ranging between 10$^5$ and 10$^7$. Applying this ratio to NGC\,4051 leads to a total gas mass inflow rate of $\ge$\,1M$_\odot$\,yr$^{-1}$.

Similar inflows in warm H$_2$ gas were observed (using again NIFS) in the inner 350\,pc of the Seyfert 2 galaxy Mrk\,1066 at 35\,pc spatial resolution \cite{riffel11} along spiral arms which seem to feed a compact rotating disk with a 70\,pc radius. The mass of warm H$_2$ gas is estimated as 3300\,M$_\odot$, that, corrected for the 10$^{5-7}$ factor to account for the cold component, would imply a reservoir of at least 10$^8$M$_\odot$ of cold molecular gas. The mass inflow rate, when corrected for this same ratio results 0.6\,M$_\odot$\,yr$^{-1}$.

Other authors have also reported inflows in near-IR observations of the H$_2$ kinematics, such as \cite{davies09} in the nuclear spirals of NGC\,1097 and \cite{muller09} between the nuclear ring and the AGN in NGC\,1068.

\section{Outflows}

Our IFS observations of the inner few hundred parsecs of nearby active galaxies have also revealed outflows. In the near-IR, while the H$_2$ gas kinematics is dominated by circular rotation or inflows in the plane of the galaxy, the ionized gas emission, and in particular in the emission lines [Fe\,II]$\lambda1.644\mu$m and [Fe\,II]$\lambda1.257\mu$m usually traces outflows   extending to high galactic latitudes. While the inflow velocities are of the order of 50\,\kms, and  the rotation velocities reach at most 200\,\kms, the outflow velocities reach up to 1000\,\kms.

One example is the case of Mrk\,1066. The NIFS observations of the inner 350\,pc (radius) show a distinct flux distributions and kinematics for the ionized gas when compared to that of the H$_2$.  The H$_2$ emitting gas is distributed all over the inner disk of the galaxy, showing circular rotation in the galaxy plane with velocities smaller than 200\,km\,s$^{-1}$. The [Fe\,II]$\lambda$1.644$\mu$m emitting gas, on the other hand, is collimated along a nuclear radio jet \cite{riffel11} and reaches outflow velocities of up to 500 \,km\,s$^{-1}$. From the velocity field and geometry of the outflow we estimate a mass outflow rate in ionized gas of  $\approx$\,0.5\,M$_\odot$\,yr$^{-1}$, a value which is of the same order as that of the mass inflow rate in H$_2$.

NIFS observations of the inner 560\,pc$\times$200\,pc of the Seyfert 1.5 galaxy NGC\,4151 at $\approx$\,7\,pc spatial resolution \cite{sb09,sb10} also show that the H$_2$ and [Fe\,II] flux distributions and kinematics are very distinct, presenting the same behavior to that observed in Mrk\,1066: the H$_2$ shows a very small velocity amplitude ($\le$\,100\,\kms) and its kinematics is dominated by rotation, while the velocities observed for [Fe\,II] reach 600\,\kms\ and the kinematics is dominated by a conical outflow. The H$_2$ gas is concentrated within the inner 50\,pc around the nucleus and is strongest along the direction of the galactic bar, consistent with an origin in an inflow along the bar. From the many [Fe\,II] and H$_2$ emission lines it was possible to obtain the electronic gas temperatures: T(H$_2$)\,$\approx$\,2000\,K, while T([Fe\,II])\,$\approx$\,15000\,K, confirming that they correspond to distinct gas components. From the observed velocities and inferred geometry, we were able to estimate the mass flow rate along the conical outflow: $\approx$\,2\,M$_\odot$\,yr$ ^{-1}$. We were also able to obtain the kinetic power of the outflow which is only $\approx$0.3\% of the bolometric luminosity of the AGN.

\begin{figure}
\includegraphics[scale=0.9]{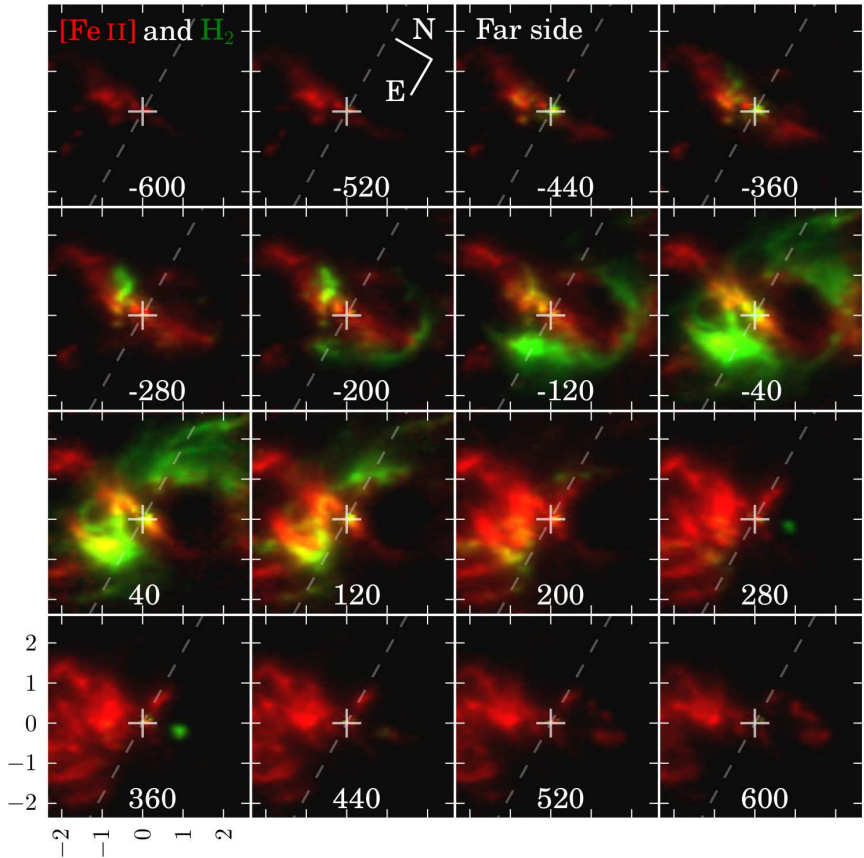}
\caption{Channel maps of the inner 200\,pc (radius) of NGC\,1068 in the [Fe\,II]$\lambda1.644\mu$m (red) and in the H$_2\lambda2.122\mu$m (green) emission lines. The numbers correspond to the central channel velocities in \kms. The nucleus is identified by a cross while the galaxy major axis is identified by a dashed line.}
\label{n1068}
\end{figure}

In Fig.\ref{n1068} we show channel maps of the inner 200\,pc radius of the ``prototypical'' Seyfert 2 galaxy NGC\,1068 in the [Fe\,II]$\lambda1.64\mu$m (in red) and H$_2\lambda2.122\mu$m (in green) emission lines, at a spatial resolution of 7\,pc \cite{barbosa13}. These channel maps show the flux distributions at the velocities indicated in each panel. The total flux distribution in [Fe\,II] emission shows an  hourglass structure, while in the channel maps of Fig.\ref{n1068} it shows an ``$\alpha$-shaped" structure in the blueshifted channels and a ``fan-shaped" structure in the redshifted channels. We attribute the blueshifted emission to the front part of the gas outflow modeled by \cite{das06}, while the redshifted emission is attributed to the back part of the outflow. We note that the fan-shaped structure is very similar to that observed in Planetary Nebulae (e.g. NGC\,6302), suggesting a similar mechanism for the origin of the outflow.  Using the inferred geometry and velocity field, we have calculate a mass outflow rate of 6\,M$_\odot$\,yr$ ^{-1}$.

Fig.\,\ref{n1068} also shows that the H$_2$ flux distribution is once more completely distinct from that of [Fe\,II], presenting a ring-like (radius\,$\approx$\,100\,pc) morphology. The H$_2$ kinematics shows again much smaller velocities than those observed in the [Fe\,II] emission and in common with other active galaxies shows also rotation. Nevertheless, something that was not seen in the other galaxies is the presence of expansion in the ring in the plane of the galaxy: Fig.\ref{n1068} shows that the H$_2$ emission is blueshifted in the near side of the galaxy and redshifted in the far side. If the gas is in the plane of the galaxy, this implies expansion of the ring. 

Additional results of IFS observations of outflows in active galaxies can be found in \cite{barbosa09,riffel11b,riffel13}.

\section{Stellar Population}

\begin{figure}
\includegraphics[scale=0.44]{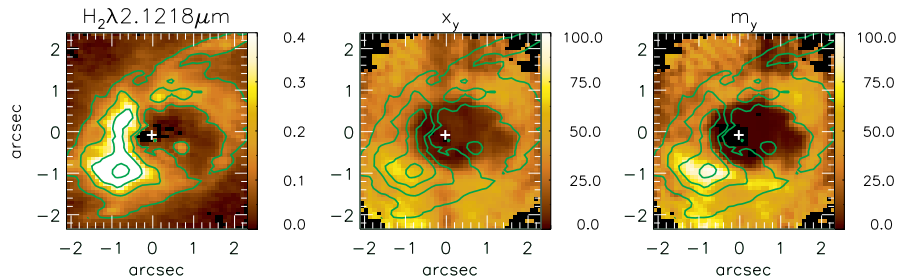}
\caption{Left panel: H$_2$ flux map of the inner 200\,pc radius of the galaxy NGC\,1068. Central panel: contours of the H$_2$ flux map overploted on the map of the percent contribution of the 30\,Myr age stellar population to the total light at 2.1$\mu$m. Right panel: percent contribution of the 30\,Myr age component to the total stellar mass. From \cite{sb12}.}
\label{pop_1068}
\end{figure}

In at least three studies with NIFS in the near-IR we have been able not only to map the flux distribution and kinematics but also the stellar population via spectral synthesis. Our study of the stellar population in Mrk\,1066 \cite{riffel11} was the first resolved two-dimensional  stellar population study of an AGN host  in the near-IR.  The spectral synthesis of the region within 350\,pc from the nucleus revealed a 300\,pc circumnuclear ring of 500\,Myr old stellar population which is correlated with a ring of low stellar velocity dispersions. This result has been interpreted as due to capture of gas to the nuclear region with enough gas mass to trigger the formation of new stars 500\,Myr ago. The low velocity dispersion indicates that these stars still keep the ``cold" kinematics of the gas from which they were formed.  In \cite{rogerio11}, a similar study revealed almost the same result for another Seyfert\,2 galaxy, Mrk\,1157. More recently, in \cite{sb12}, we have found a smaller (100\,pc radius) ring of star formation, with a younger age, of $\approx$\,30\,Myr, in NGC\,1068. This ring seems to be correlated with the molecular (H$_2$) ring described above, and shown in Fig.\,\ref{pop_1068} (in green).

In \cite{barbosa06}, in a study of the gas and stellar kinematics of the inner kiloparsec of six nearby Seyfert galaxies using the GMOS IFU, we have also found that four of them showed circumnuclear rings of low velocity dispersion, supporting the presence of young to intermediate age stars with ``cold" kinematics.

\section{Summary and Conclusions}

Our observations of the gas kinematics within the inner few hundred parsecs of nearby active galaxies have revealed inflows towards the center along nuclear spirals, with inflow velocities in the range  50--100\,\kms and mass flow rates ranging from 0.01 to 1\,M$_\odot$\,yr$ ^{-1}$ in ionized gas via observations in the optical. In the near-IR, we have also found inflow along nuclear spirals in warm molecular gas emission (T$\sim$\,2000\,K), although at  much smaller inflow rates. We have concluded that the warm molecular gas is only the hot ``skin" of the bulk inflow that should be dominated by cold molecular gas (observable with ALMA). Using a typical conversion factor between the mass of warm to cold molecular gas (derived from an active galactic sample for which both millimetric and near-IR observations are available), we find values close to 1\,M$_\odot$\,yr$ ^{-1}$.

Our previous study \cite{sl07} showed a clear dichotomy between the nuclear region of early-type AGN hosts -- which always show excess of dust -- and that of non-AGN, supporting the hypothesis that the nuclear spirals and filaments are a necessary condition for the presence of nuclear activity. Nevertheless, the mass accretion rate necessary to feed the AGN ($\approx$\,10$^{-3}$\,M$_\odot$\,yr$ ^{-1}$) is typically much smaller than the above mass inflow rate.  For example, at 1\,M$_\odot$\,yr$ ^{-1}$, in 10$^{7-8}$\,yr (the expected duration of an activity cycle), at least 10$^7$M$_\odot$ of gas will be accumulated in the inner few hundred parsecs. This number is supported by our recent study \cite{martini13} in which we have used Spitzer photometry of  the sample of \cite{sl07} to obtain dust masses. The average value for the AGN hosts is 10$^{5.5}$M$_\odot$, and for a typical ratio of $\approx$\,100 between the gas and dust masses, the AGN hosts should thus have, on average, 10$^{7.5}$M$_\odot$ in gas in the inner few hundred parsecs, in agreement with our estimate above on the basis of the mass inflow rate.

The accumulation of at least 10$^7$M$_\odot$ in the nuclear region will probably lead to the formation of new stars in the galaxy bulge. Signatures of this recent star formation have been indeed seen, in the form of rings at 100\,pc scales with significant contribution from stars of ages in the range 30\,Myr$\leq$age$\leq$\,700\,Myr. These results suggest that we are witnessing the co-evolution of the SMBH and their host galaxies in the near Universe: while the SMBH at the center grows at typical rates of 10$^{-3}$\,M$_\odot$\,yr$ ^{-1}$, the bulge growths at typical rates of 0.1--1\,M$_\odot$\,yr$ ^{-1}$. A similar evolution scenario has been previously proposed by \cite{sb00}.

Finally, the coupling between the SMBH and galaxy evolution depends also on the AGN feedback. Our IFS observations usually reveal ionized gas outflows in the nuclear region, mainly in Seyfert hosts. These outflows are oriented at random angles to the galaxy plane, and reach velocities in the range 200--800\,\kms. The total mass of ionized gas in these outflows are of the order of 10$^{6-7}$\,M$_\odot$, and the mass outflow rates are in the range 0.5--10\,M$_\odot$\,yr$ ^{-1}$, which is similar to the range of the mass inflow rates, although only in a couple of galaxies we observe both inflows and outflows (e.g. in Mrk\,1066). The fact that the mass outflow rates are about 1000 times the AGN accretion rate supports the idea that the observed outflows are due to mass loading of an AGN outflow (which should be at most equal to the AGN accretion rate) as it moves through the circumnuclear interstellar medium of the host galaxy.
 
{}   
\clearpage
\end{document}